# The evolution of habitable zones during stellar lifetimes and its implications on the search for extraterrestrial life

David R. Underwood, Barrie W. Jones, P. Nick Sleep
*Dept of Physics & Astronomy, The Open University, Milton Keynes, MK7 6AA, UK*
*e-mail: d.r.underwood@open.ac.uk*

**Abstract:** A stellar evolution computer model has been used to determine changes in the luminosity $L$ and effective temperature $T_e$ of single stars during their time on the main sequence. The range of stellar masses investigated was from 0.5 to 1.5 times that of the Sun, each with a mass fraction of metals (metallicity, $Z$) from 0.008 to 0.05. The extent of each star's habitable zone (HZ) has been determined from its values of $L$ and $T_e$. These stars form a reference framework for other main sequence stars. All of the 104 main sequence stars known to have one or more giant planets have been matched to their nearest stellar counterpart in the framework, in terms of mass and metallicity, hence closely approximating their HZ limits. The limits of HZ, for each of these stars, have been compared to its giant planet(s)'s range of strong gravitational influence. This allows a quick assessment as to whether Earth-mass planets could exist in stable orbits within the HZ of such systems, both presently and at any time during the star's main sequence lifetime. A determination can also be made as to the possible existence of life-bearing satellites of giant planets, which orbit within HZs. Results show that about half of the 104 known extrasolar planetary systems could possibly have been housing an Earth-mass planet in HZs during at least the past billion years, and about three-quarters of the 104 could do so for at least a billion years at some time during their main sequence lives. Whether such Earth-mass planets could have formed is an urgent question now being investigated by others, with encouraging results.

**Introduction**

The search for the signs of carbon-based extraterrestrial life outside our Solar System is expected to focus on extrasolar planets within the habitable zones of exoplanetary systems around stars similar to the Sun. The habitable zone (HZ) around a star is defined as the range of distances over which liquid water could exist on the surface of a terrestrial planet, given a dense enough atmosphere (Jones *et al*, 2001). The planet would be rocky, be of order of the Earth's mass, and would have had to lie in an orbit within the HZ for at least the order of a billion years (= 1 Gy) for life to have had a detectable effect on its atmosphere. This duration is based on the effect of Earth's biosphere on our atmosphere (Chyba, 1993).

The detection of Earth-mass extrasolar planets is, at the moment, beyond most current technology. This is shortly due to be redressed, however, with the launch of planet detecting satellites such as the transit-detecting Kepler mission (2008) and the direct imaging missions Darwin (ESA) and TPF (NASA) (2015), plus imaging by extremely large 30-100 m telescopes with high performance adaptive optics (e.g. ESO's OWL). We can, meanwhile, use computer models to predict in which of the currently known exoplanetary systems Earth-mass planets could exist, where 119 extrasolar planets of the order of Jupiter's mass have been found in orbit around 104 stars (catalogued by Jean Schneider at http://www.obspm.fr/encycl/encycl.html). Existence requires that the giant planets would allow terrestrial bodies to remain in **stable** orbits around their parent star i.e. in orbits that do not result in collision or ejection. To determine this, a mixed variable symplectic computer integrator has been used previously (Jones & Sleep, 2002; Jones et al., 2001), to simulate the orbital motion of Earth-mass planets over one billion years, in a sample of such systems. If a terrestrial planet could exist in a stable orbit for this length of time it is likely to be able to exist there for the duration of that star's main sequence lifetime.

A putative terrestrial planet may exist in an orbit that is not only stable but remains **confined** to its star's HZ for some period i.e. the semimajor axis remains in the HZ even if the eccentricity carries it outside the HZ for part of each orbit. If confinement could last for the whole of the main sequence lifetime, the planet would be in the continuously habitable zone and the system would be a very good candidate in searching for signs of extraterrestrial life. Alternatively, a planet may only remain within the HZ for part of a star's main sequence lifetime, due to the outward migration of the HZ as the star ages. But if this period includes the last billion years the system would still be a good candidate for the detection of life, provided that the period excluded an epoch of heavy bombardment soon after planetary formation, such as afflicted the Earth for its first 0.7 Gy. We have established to which of these categories each of the known exoplanetary systems belongs.

In order to map the outward movement of the HZs around these known exoplanetary systems during their stars' main sequence lifetimes, our first step was to model the main sequence lifetime of theoretical stars. This allowed us to follow the changes in their total power output i.e. the luminosity $L$, and in the surface temperature as represented by the effective temperature $T_e$, these being the parameters on which the location of the HZ depends. The stellar evolution program continued into the first period of the giant phase that follows the main sequence, however this period was disregarded as complications arise due to stars losing mass during these stages, and in any case the transition to giant would sterilise a planetary system. A range of masses, 0.5 to 1.5 times that of the Sun, and metallicities $Z$ (mass proportion of elements other than hydrogen and helium) from 0.008 to 0.05, were modelled, 55 combinations in total. This created a framework of F, G, K and early M type main sequence stars. Such stars are more abundant than the more massive O, B and A stars, and have main sequence lifetimes long enough for life to have had an observable effect on exoplanetary atmospheres. Then, for each framework star, the migration of the HZ was determined over its main sequence lifetime. Stars known to have exoplanetary systems are then compared to their nearest matching stellar type within this framework. The ranges of their giant planet(s)'s orbital motion and gravitational influence are mapped onto the HZ plots. This enables a rapid assessment as to whether terrestrial planets could have been confined to the HZs of such systems for at least a billion years before the present. A 'rating value' is then attributed, which is a measure of the chance of the planetary system supporting life. It is hoped this information will help direct search efforts to the exoplanetary systems most likely to have detectable life.

Though at present the number of exoplanetary systems is only about twice the number of framework stars (55), many more exoplanetary systems are expected to be discovered in the next few years, in which case the use of a framework will provide an increasingly significant saving of time.

## Creating the framework of main sequence stars' luminosities and effective temperatures

Computer simulations of stellar main sequence evolution were carried out using a model developed by Mazzitelli (1989). The stellar input parameters that can be varied are shown in Table 1. The mixing length ($\alpha$) is the ratio of the distance a rising gas bubble travels before equilibrating with its surroundings, to the pressure scale height $p/\rho g$, where $p$ is pressure, $\rho$ is density and $g$ is acceleration due to gravity.

Table 1. *Stellar parameters that can be varied*

| Symbol | Name | Definition |
|---|---|---|
| $Y$ | helium content | Mass fraction of helium in star at $t = 0$ |
| $Z$ | metallicity | Mass fraction of elements other than H and He in star at $t = 0$ |
| DEFAUC | carbon content | Mass fraction of carbon in $Z$ at $t = 0$ |
| DEFAUN | nitrogen content | Mass fraction of nitrogen in $Z$ at $t = 0$ |
| DEFAUO | oxygen content | Mass fraction of oxygen in $Z$ at $t = 0$ |
| ALFA ($\alpha$) | mixing length | Distance a rising gas bubble travels before equilibrating with its surroundings divided by the pressure scale height ($p/\rho g$) |
| | | *At zero age main sequence (ZAMS) ...* |
| EMMU | initial mass | Stellar mass |
| ELLU | initial luminosity | Stellar luminosity |
| ELLTE | initial effective temperature | Stellar effective temperature |
| ELLPC | initial core pressure | Stellar core pressure |
| ELLTC | initial core temperature | Stellar core temperature |

A simulation starts from the zero age main sequence position of the star i.e. when nuclear reactions commence to burn hydrogen in its core. Each parameter in Table 1 has to be specified, though the values of ELLU, ELLTE, ELLPC, and ELLTC only have to be plausible – the simulation will then produce values appropriate to the stellar model a short way into the simulation.

Evolutionary trends were monitored by parameter outputs in maximum time steps of a million years, until the end of the main sequence. (The program then proceeded into post main sequence phases with time steps adjusted automatically to allow for rapid stellar changes.) At the end of each run, the information generated was imported onto a spreadsheet. The luminosity $L$ and effective temperature $T_e$ enable us to calculate HZ boundaries, as described in the next section. We also recorded stellar radius $R$, which assists in calibrating the model, to which we now turn.

Calibration used the present day Sun as the standard. This involved altering the mass proportions of helium ($Y$), carbon (DEFAUC), nitrogen (DEFAUN), oxygen (DEFAUO), metallicity ($Z$), and the mixing length ($\alpha$), until the Sun was at its present day $L$, $T_e$ and $R$, at the correct age. The values of $Y$, DEFAUC, DEFAUN, DEFAUO, and $\alpha$, were then used in all subsequent stellar runs. $Z$ was varied to match the particular star, as (of course) was the initial mass, EMMU.

The present day $L$, $T_e$ and $R$ for the Sun were reproduced with the values $\alpha = 1.9$, $Y = 0.269$, DEFAUC = 0.1516, DEFAUN = 0.05289, DEFAUO = 0.52889, and $Z = 0.02$. These parameters gave $L = 1.000\ L_\odot$, $T_e = 5787$ K, $R = 0.995\ R_\odot$, at an age of 4.578 Gy, where $L_\odot$ is the present day solar luminosity and $R_\odot$ is the current solar radius. The actual present day values, at 4.55 Gy, are $T_e = 5800$ K, and (of course) $L_\odot$ and $R_\odot$.

The solar values from the simulation are in good agreement with current measurements and occur at essentially the correct age. All of these parameters are in acceptable ranges. The solar main sequence profile, derived from this run, is shown in Fig. 1. Note that this extends a little way beyond the main sequence phase, which ends at about 11 Gy.

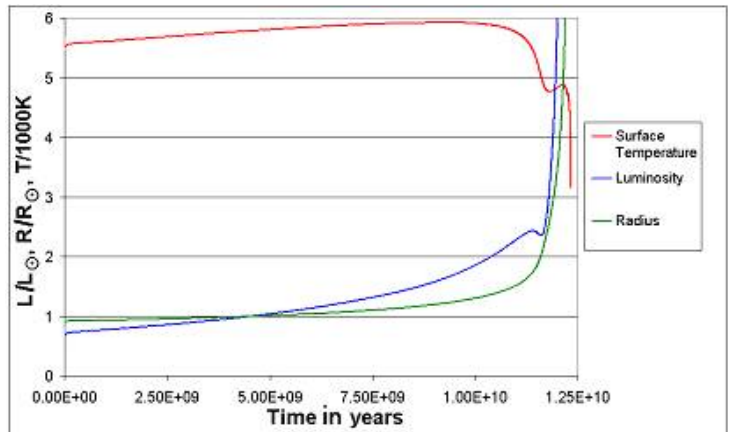

**Fig. 1** The Sun's main sequence effective temperature, luminosity and radius

For the simulation of other stars, the mass proportions of helium, carbon, nitrogen, and oxygen, and the mixing length, were set the same as those for the Sun. These mass proportions are unlikely to be significantly different from the Sun among nearly all the stars studied, nearly all of which are within a few hundred light years of the Sun. Although the mixing length might not be 1.9 for all the stellar masses modelled, variations cause only small differences to simulations, and primarily to the star's radius. Individually set were the stellar mass and metallicity, and the initial values of luminosity, effective temperature, core temperature and core pressure

The framework stars are in the range 0.5 to 1.5 solar masses, separated at 0.1 solar mass intervals, covering spectral types from early M through K, G and F. At each mass, metallicities of 0.008, 0.013, 0.02, 0.032 and 0.05 were used. This spread of masses covers main sequence lifetimes long enough for any biospheres to have had an observable effect on planet atmospheres. The ranges adequately cover the masses and metallicities of stars known to have extrasolar planetary systems (catalogued by Jean Schneider at http://www.obspm.fr/encycl/encycl.html). Figure 2 shows the framework values. (Note that for the metallicity of 0.05, the masses at 1.2 and 1.3 solar masses were adjusted to 1.15 and 1.35 solar masses respectively to facilitate program running.)

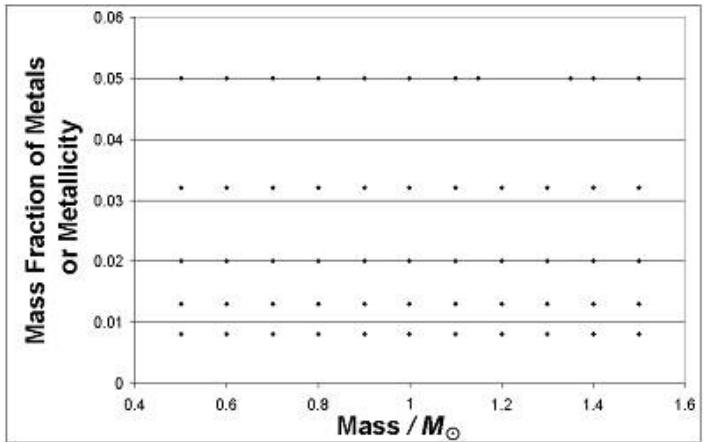

**Fig. 2** Framework values of stellar mass and metallicity

For models with the same mass, decreasing their metallicity decreases the main sequence lifetime and increases the luminosity and effective temperature, although this effect is not as great as an increase in stellar mass. This is illustrated in Figure 3, which should be compared with the solar case in Figure 1.

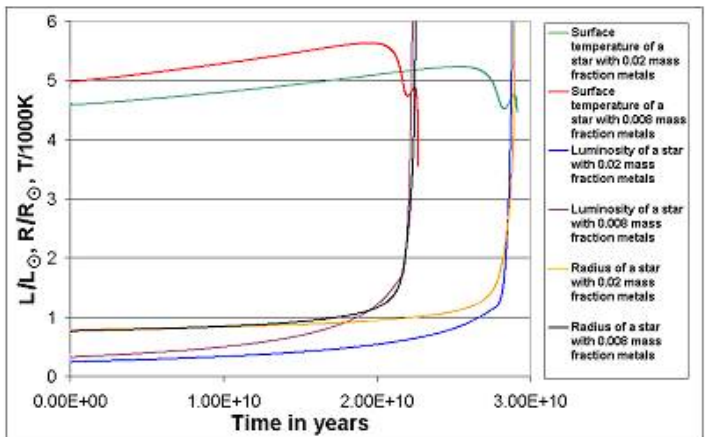

**Fig. 3** Luminosity, effective temperature and radius of a star with a mass of 0.8 solar masses, and with a solar metallicity plus a lower metallicity

For a few framework stars we identified actual stars with similar masses and metallicities, and with age estimates. The measured luminosities and effective temperatures of these stars were similar to those from the model, though age estimates are insufficiently precise for this to be a severe test. The next task was to establish the HZ boundaries for the stars known to have exoplanetary systems.

**Habitable zone boundaries**

The HZ evolves during a star's main sequence lifetime – the boundaries generally migrate outwards. During this time, nuclear burning of hydrogen builds up a helium residue in a stellar core causing an increase in pressure and temperature. This occurs more rapidly in stars that are more massive and lower in metallicity. These changes are transmitted to the outer regions of a star, which results in a steady increase in luminosity and changes in effective temperature. The major effect is that a star becomes more luminous, causing HZ to move outwards.

The change in a star's effective temperature will have a smaller yet still noticeable effect on the HZ. A redder star, with a lower effective temperature, will have a spectrum weighted more towards the infrared, as determined by Wien's Law:

$$\lambda_{peak} = k/T \qquad (1)$$

where $\lambda_{peak}$ is the peak wavelength of the emission profile, $T$ is absolute temperature and $k$ is a constant. For two stars with the same luminosity, the redder will have its HZ at a greater distance because the greater proportion at infrared wavelengths generally makes the star more effective in raising the temperature of the surface of a planet. A star's effective temperature increases during most of the main sequence, so the luminosity increase causing the HZ to move out will be slightly offset. However, during the later main sequence phase, when its surface begins to cool as it approaches its red giant phase, the outward movement of the HZ will now be enhanced. This movement could result in a life-bearing planet, within the HZ at the beginning of a star's main sequence lifetime, becoming too hot and hence uninhabitable. Similarly, a lifeless planet originally outside the HZ, may thaw out and enable life to commence. Figure 4 illustrates the positional change of the Sun's HZ with time, with the inner boundary calculated for a runaway greenhouse criterion and the outer boundary calculated for a maximum greenhouse criterion, both of which are explained shortly.

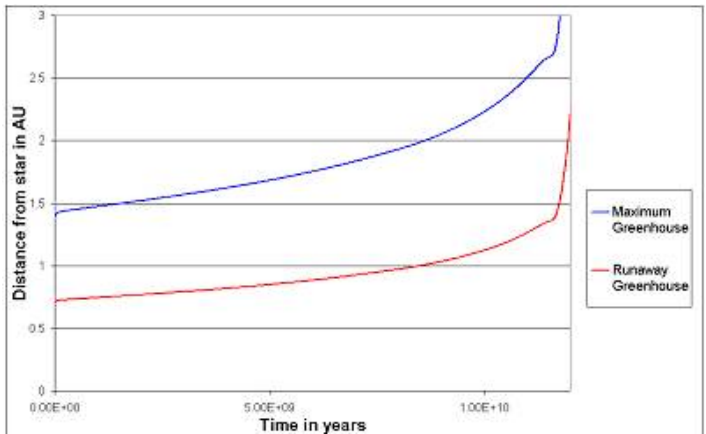

**Fig. 4** The Sun's habitable zone

The duration for which a planet remains in the HZ depends on the width of the HZ and on how fast it migrates during the star's main sequence lifetime. These factors depend on the atmospheric model used to define the HZ limits and on the precise criteria adopted. For one well-known and still-used atmospheric model three criteria have been applied to determine the inner boundary of the HZ, and three to define the outer boundary (Kasting *et al.*, 1993). At the inner boundary the criteria are, in order of decreasing stellar flux:

1. **Recent Venus**  Assuming Venus formed with water, the solar flux at Venus $10^9$ years ago, when water is believed to have last flowed (Solomon & Head, 1991).
2. **Runaway greenhouse effect**  Water vapour enhances the greenhouse effect and thus promotes surface warming. This increases the atmospheric vapour content, thus further raising the surface temperature. At sufficient stellar flux this will lead to the rapid evaporation of all surface water. The temperature at the surface of the planet reaches the critical point of water, 647 K. Water is soon lost from the upper atmosphere by photodissociation and subsequent hydrogen loss to space.

3. **Water loss** This occurs in an atmosphere warm enough to have a wet stratosphere, from where water is gradually lost by photodissociation and subsequent hydrogen loss to space.

At the outer boundary the criteria are, in order of decreasing stellar flux:
4. **First carbon dioxide condensation**. For a surface temperature of 273K, the distance at which carbon dioxide clouds first begins to form.
5. **Maximum greenhouse effect**. The maximum distance at which a surface temperature of 273 K can be maintained by a cloud-free carbon dioxide atmosphere.
6. **Early Mars**. The flux at Mars at the beginning of the solar main sequence lifetime, when free standing bodies of water are believed to have existed.

Criteria 1 and 6 define the limits of the broadest HZ, when life would have the longest time to evolve on a planet. The narrowest HZ would be for criteria 3 and 4. However, in 2 and 3 the cloud cover is fixed at the present terrestrial value – too little is known about cloud formation to do better than this. An increase in water clouds would have the effect of increasing a planet's albedo, hence cooling the planet and moving the inner boundary of the HZ inwards. At the outer edge, under criterion 5, the formation of carbon dioxide clouds could have a net warming effect, through a scattering greenhouse effect (Forget and Pierrehumbert, 1997). This would move the outer boundary of the HZ outwards. We will, however, use the criteria unmodified.

**Habitable zone boundaries of stars known to have planetary systems**

Kasting et al. (1993) have published stellar fluxes corresponding to the three pairs of HZ boundary criteria above, for three effective temperatures. These are 3700 K, 5700 K, and 7200 K, corresponding to typical M0, G2 (= Sun), and F0 main sequence stars. These data are shown in Table 2, where $S$ is the stellar flux at the HZ boundary in units of the solar constant, 1368 W m$^{-2}$. Also shown are the luminosities $L$ that are paired with $T_e$, and the corresponding distances $d$ of the HZ boundaries from the star, using the inverse square law i.e.
$$d = (L/S)^{1/2} \qquad (2)$$
The luminosity value of the F0 star has been updated from the value used by Kasting et al. (Cox, 2000).

Table 2. *Critical stellar fluxes and HZ boundary distances for various stars*

| Stellar type (main sequence) | M0 | | G2 | | F0 | |
|---|---|---|---|---|---|---|
| Effective temperature $T_e$/K | 3700 | | 5700 | | 7200 | |
| Luminosity $L/L_\odot$ | 0.06 | | 1.00 | | 6.45 | |
| HZ Boundary | $S$* | $d$/AU | $S$* | $d$/AU | $S$* | $d$/AU |
| Recent Venus** | 1.60 | 0.19 | 1.76 | 0.75 | 2.00 | 1.80 |
| Runaway greenhouse | 1.05 | 0.24 | 1.41 | 0.84 | 1.90 | 1.85 |
| Water loss | 1.00 | 0.25 | 1.10 | 0.95 | 1.25 | 2.28 |
| First CO$_2$ condensation | 0.46 | 0.36 | 0.53 | 1.37 | 0.61 | 3.26 |
| Maximum greenhouse | 0.27 | 0.47 | 0.36 | 1.67 | 0.46 | 3.76 |
| Early Mars*** | 0.24 | 0.50 | 0.32 | 1.77 | 0.41 | 3.98 |

\* The values are relative to the present mean solar flux at the top of Earth's atmosphere (1368 W m$^{-2}$).

| * * | Recent Venus fluxes, for both M0 and F0 stars, are derived from their water loss fluxes multiplied by the ratio of the flux values of recent Venus over water loss of the G2 star. |
| *** | Early Mars fluxes, for both M0 and F0 stars, are derived from their maximum greenhouse fluxes multiplied by the ratio of the flux values of early Mars over maximum greenhouse for the G2 star. |

The core data in Table 2 are the relationships between $S$ and $T_e$. The three values for each HZ criterion can be uniquely fitted with a parabolic equation, to give the following outcome (with $T_e$ in K).

| | |
|---|---|
| Recent Venus | $S = 2.286 \times 10^{-8} T_e^2 - 1.349 \times 10^{-4} T_e + 1.786$ |
| Runaway greenhouse | $S = 4.190 \times 10^{-8} T_e^2 - 2.139 \times 10^{-4} T_e + 1.268$ |
| Water loss | $S = 1.429 \times 10^{-8} T_e^2 - 8.429 \times 10^{-5} T_e + 1.116$ |
| First $CO_2$ condensation | $S = 5.238 \times 10^{-9} T_e^2 - 1.424 \times 10^{-5} T_e + 0.4410$ |
| Maximum greenhouse | $S = 6.190 \times 10^{-9} T_e^2 - 1.319 \times 10^{-5} T_e + 0.2341$ |
| Early Mars | $S = 5.714 \times 10^{-9} T_e^2 - 1.371 \times 10^{-5} T_e + 0.2125$ |

These equations can be used to generate $S$ for any value of $T_e$ from our stellar evolution model. The luminosity $L$ from the model can then be used in equation (2) to calculate the values of $d$ for the HZ boundaries. In particular, this is done for the stars known to have exoplanetary systems. Each of these stars has a known mass and metallicity, and in most cases there is an estimate of the age (catalogued by Jean Schneider at http://www.obspm.fr/encycl/encycl.html). The framework star nearest to the actual star is identified and the values of $d$ for the HZ boundaries throughout the main sequence lifetime are calculated. The framework is sufficiently fine for this to be an adequate procedure.

Fig. 5 shows the HZ boundaries for two very different stars with planets, for all six boundary criteria. The slow outward movement of HZ in the low mass (0.8$M_\odot$) star Epsilon Eridani, Fig. 5a, show that a planet in a stable orbit would remain in the HZ for more than 25 Gy. In contrast, HZ moves outward more rapidly for the more massive (1.25$M_\odot$) Tau[1] Gruis, Fig. 5b, and a planet would remain in the HZ for only 5 Gy.

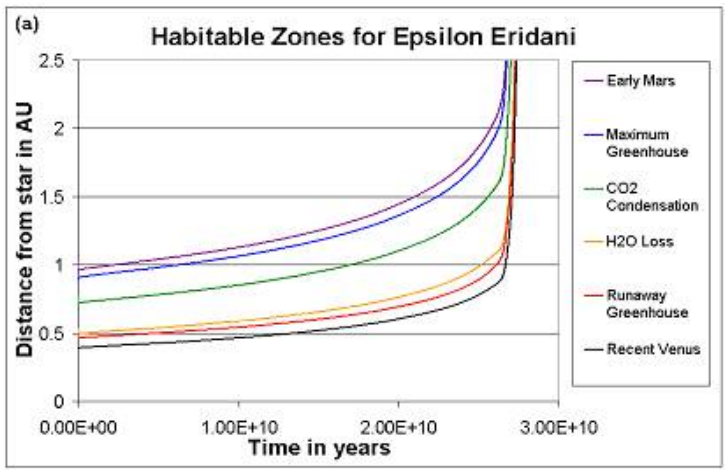

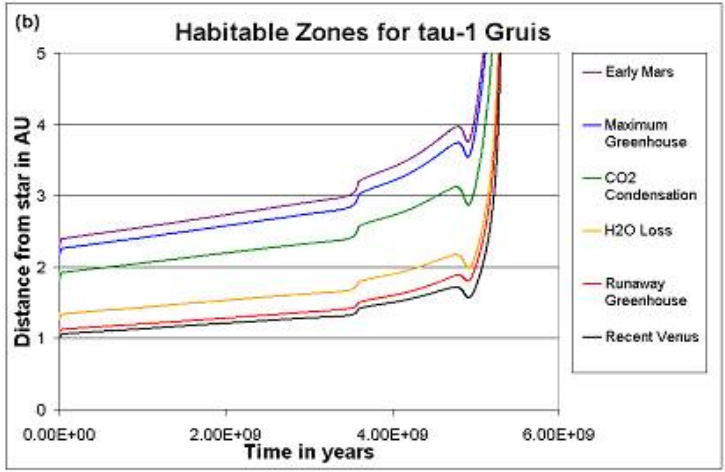

**Fig. 5** Two examples of HZ boundaries versus time showing all six boundaries

The boundaries in Fig. 5 have been derived by us from the atmospheric models of Kasting et al. (1993), but with updated stellar evolution models. More recent work, with other models, suggests that the runaway greenhouse – criterion 2 – would occur only somewhat closer to the star (Williams and Pollard, 2002). Work by Williams and Kasting (1997), Forget and Pierrehumbert (1997), and Mischna et al. (2000), indicates that the inner boundary for the first $CO_2$ condensation – criterion 4 – could be somewhat further from the star, as could the maximum greenhouse – criterion 5 – if cloud formation occurs.

In the remainder of this work we have adopted the intermediate criteria (2 and 5) with the Kasting 1993 atmospheric model. Bearing in mind that, on these criteria, the inner boundary could be somewhat closer in, and the outer boundary could be somewhat further out, this has resulted in a rather conservative view of which exoplanetary systems could harbour Earth-mass planets in the HZ.

**Categorisation of exoplanetary systems by detectability of life**

It was argued above that if a terrestrial planet could have been confined to the HZ for at least the past billion years it might be possible to detect the presence of a biosphere there today. To make this assessment we apply a criterion that we have developed through the detailed study by numerical integration of several exoplanetary systems (Jones et al., 2004). According to this criterion, if a planet with a mass the order of that of the Earth is launched with a semimajor axis within a certain distance of a giant planet in the system, then confinement will probably be for less than a billion years, usually considerably less. If it is launched outside this distance it will probably be confined for the whole main sequence lifetime. The critical distance $D$ from the giant is given by

$$D = nR_H \qquad (3)$$

where $n$ is in the approximate range 3-8 (see below), and $R_H$ is the Hill radius of the giant planet defined by

$$R_H = a_G (m_G/3M)^{1/3} \qquad (4)$$

where $a_G$ is the semimajor axis of the giant planet's orbit, $m_G$ is the giant's mass, and $M$ is the mass of the star. A small body at $R_H$ will experience roughly equal gravitational influences from the giant planet and star.

At around $D$ there is a decline in the confinement time from the full term of the integration (typically 1 Gy of simulated time), to a small fraction of this, the 'Earth' being ejected from the planetary system, or colliding with the star or the giant planet. We have found (Jones et at, 2004) that $n$ is a little less than 3 for an Earth-mass planet with a semimajor axis less than the giant's. For Earth-mass planets with larger semimajor axes than the giant, $3 \leq n \leq 8$, where $n \approx 3$ when the giant has a near circular orbit and $n$ approaches 8 when the giant has an eccentric orbit.

For the discovered systems, the planet masses are minimum values, corresponding to the *assumption* that we are seeing the orbits edgewise – the Doppler spectroscopy method that has given us almost all the information we have on exoplanets does not reveal the angle of view. The planetary masses are thus lower limits. Planets with orbits inclined to the line of sight will have larger true masses, however equation (4) shows that $R_H$ is proportional to $m_G^{1/3}$. A planet with twice its minimum mass would orbit in a plane tilted at the high inclination of $60^O$ to Earth's line of sight, yet this would increase $R_H$ by a factor of only 1.26. Now for Earth-mass planets with smaller semimajor axes and orbiting inside the giant, $n = 2.5$-$3.0$. So, in the majority of cases, even quite highly inclined orbits would not increase $n$ above about 3. The value of $n$ for Earth-mass planets with larger semimajor axes and orbiting outside the giant depends on the giant planet's orbital eccentricity, and the values quoted above, where $3 \leq n \leq 8$ again apply well enough to quite high inclinations.

Fig. 6 shows four different types of exoplanetary system representative of those discovered so far. The four stars have similar masses of $0.9\ M_\odot - 0.95\ M_\odot$, hence similar main sequence lifetimes and illustrate how each exoplanetary system is assessed for potentially habitable 'Earths', as explained in the extensive Fig. 6 caption. Recall that the HZs correspond to the intermediate boundary criteria, i.e. a runaway greenhouse effect at the inner boundary and a maximum greenhouse effect at the outer boundary. The extent of the giant planet's orbit, from apastron (furthest from the star) to periastron (closest to the star), is marked. Also shown are the critical distances $nR_H$ from the peri/apastra, within which stable orbits of Earth-mass planets are not possible.

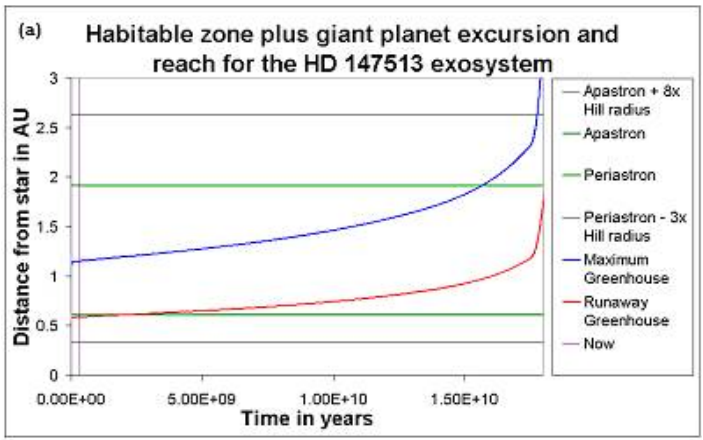

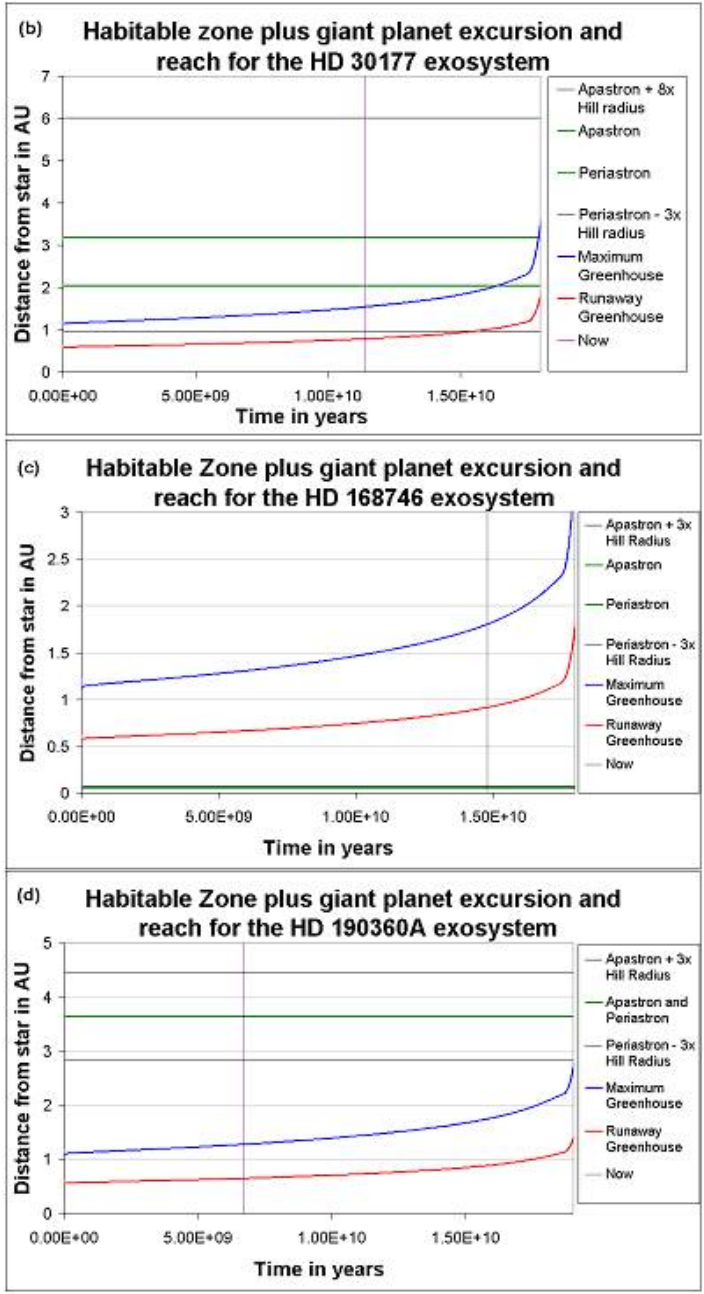

**Fig. 6** Planet and HZ configurations of four exoplanetary systems. (a) Figure 6(a) shows the HZ of HD147513, which lies closer than $nR_H$ to the giant planet throughout the main sequence lifetime, and so a life bearing Earth-mass planet could never be confined here. This giant has an orbital eccentricity $e_G = 0.52$, and so $n \sim 8$ at the outer HZ boundary. (b) Figure 6(b) shows that the inner part of the HZ of HD30177 is more than 3 $R_H$ from the giant's periastron throughout the past, but will not be so for long into the future. (c) Figure 6(c) shows the configuration for the HD 168746 system, which has a 'hot Jupiter' – a giant close to the star. Here the HZ is completely clear for an Earth-mass planet to exist in a confined and stable orbit. As the giant is so close to the star, the value of $n$ in equation (3) will have little effect on the outer gravitational reach of the giant. (d) Figure 6(d) shows the system, HD190360A, that, among the four here, is most like the Solar System, with the giant planet in a circular orbit at 3.7 AU, well beyond the HZ. Here the star is 6.7 Gy old and an Earth-mass planet could quite possibly exist in a stable orbit confined to the HZ. If an 'Earth' is at 1AU, it would have already spent its entire time within the HZ, and it would have more than 10 Gy left within the HZ for any life to continue.
------------------------(end of caption)------------------------

We have worked through all of the currently known main sequence exoplanetary systems and examined them by this method (catalogued by Jean Schneider at http://www.obspm.fr/encycl/encycl.html). The outcome is shown in Table 3. The 'now' column indicates whether an Earth-mass planet could have been confined to the HZ for at least the past billion years (excluding any initial 0.7 Gy heavy bombardment), and the 'sometime' column whether this could be the case at any time during the star's main sequence lifetime. 'Part' indicates a small proportion of the HZ offers confinement, for example near its outer boundary for a giant planet not much closer to the star than the inner boundary. A '?' in the 'now' column indicate the star's age is unknown, where this is critical to the evaluation, although even 'known' stellar ages are susceptible to errors of up to ± 20% of the star's age, except for stars younger than ~1 Gy, where the ages are better known. Information in *italics* indicates planetary systems with more than one giant planet.

Table 3 reveals promising statistics for the possibility that Earth-mass planets, and therefore life-bearing planets, exist within the HZs of known exoplanetary systems. About half of the 104 systems could have housed an 'Earth' in their HZs for at least the past billion years, and about three-quarters at sometime during their main sequence lifetime. Not included in the list of 'possibles' are putative Earth-mass satellites of giant planets in the HZs.

Although the outcome seems promising for the existence of extrasolar 'Earths', the question is begged of whether these 'Earths' could have formed at positions that would make them promising candidates for detectable life in the first place. There has been some work on this problem (Barbieri et al., 2002; Kokubo & Ida, 2002; Lissauer, 1998) but more is needed. Recently, Mandell & Sigurdsson (2003) have performed numerical simulations of the survival of terrestrial bodies during giant planet inward migration, suggesting that a significant fraction may survive this process. The systems that are probably the best candidates for having Earth-like planets in stable orbits in their HZs are those that look like the Solar System, highlighted by '**' in Table 3, since they avoid any problems associated with hot Jupiter migration. More of these systems will be discovered in the near future by the radial velocity surveys, because, by definition, they have long period orbits and the surveys are only now beginning to have acquired data for sufficiently long.

**Conclusions and future work**

A method has been successfully devised which rapidly determines the HZs of stars between $0.5M_\odot$ and $1.5M_\odot$ with metallicities between 0.008 and 0.05. Overlaying the orbital excursion of giant planets and their gravitational reach in known exoplanetary systems has enabled each system to be evaluated on its likelihood of housing a habitable terrestrial planet within the HZ. This method can now be applied to any exoplanetary system as it is discovered, to similarly determine whether it could house habitable terrestrial planets. It can also be used to re-evaluate known systems, when orbital parameters of any giant planet(s) are updated.

All of these planetary systems may be rated as to whether they are good, fair or poor candidates in the search for habitable 'Earths'. Such ratings will assist in ensuring observational searches are concentrated on the correct choice of targets. Of the known exoplanetary systems a possible life-bearing 'Earth' could presently exist in about half of the 104 systems and in about three-quarters of these systems at some time during their main sequence lifetime. This method may also be used to determine whether giant planets, with orbits of low eccentricity within the habitable zone, are able to house habitable Earth-mass satellites.

We have not, however, studied whether terrestrial planets or satellites of giant planets are able to form in the habitable zones of these systems. This will be investigated in the next stage of this research additional to investigations already undertaken by Mandell and Sigurdsson (2003).

**Acknowledgements**

We are grateful to John Barker for assistance in running the Mazzitelli stellar evolution model, and to Ulrich Kolb for discussions on stellar evolution.

**Figure Captions**

Fig. 1   The Sun's main sequence surface temperature, luminosity and radius
Fig. 2   Framework values of stellar mass and metallicity
Fig. 3   Luminosity, effective temperature and radius of a star with a mass of 0.8 solar masses, and with a solar metallicity plus a lower metallicity
Fig. 4   The Sun's habitable zone
Fig. 5   Two examples of HZ boundaries versus time showing all six boundaries
Fig. 6   Planet and HZ configurations of four exoplanetary systems

Table 3. *Classification of Habitability of Exoplanetary Systems*

| Star | now | sometime | Star | now | sometime | Star | now | sometime |
|---|---|---|---|---|---|---|---|---|
| OG-TR-56 | yes | yes | HD80606 | NO | part | HD222582 | NO | NO |
| HD73256 | yes | yes | HD219542B | yes | yes | HD65216 | ? | NO |
| HD83443 | yes | yes | 70 Vir | ? | yes | *HD160691* | *NO* | *NO* |
| HD46375 | yes | yes | HD216770 | ? | yes | HD141937 | NO | NO |
| HD179949 | yes | yes | HD52265 | yes | yes | HD41004A | part | part |
| HD187123 | yes | yes | GJ3021 | NO | part | HD47536 | NO | NO |
| Tau Boo | yes | yes | *HD37124* | *NO* | *NO* | HD23079 | NO | part |
| BD103166 | yes | yes | HD 219449 | NO | part | 16 CygB | NO | NO |
| HD75289 | yes | yes | HD73526 | part | part | HD4208 | part | part |
| HD209458 | yes | yes | HD104985 | ? | yes | HD114386 | yes** | yes** |
| HD76700 | yes | yes | *HD82943* | *NO* | *part* | gam Ceph | part | part |
| 51 Peg | yes | yes | *HD169830* | *part* | *part* | HD213240 | NO | NO |
| *Ups And* | *NO* | *NO* | HD8574 | part | yes | HD10647 | NO | part |
| HD49674 | yes | yes | HD89744 | part | yes | HD10697 | NO | NO |
| HD68988 | yes | yes | HD134987 | part | yes | *47 UMa* | *part* | *part* |
| HD168746 | yes | yes | HD40979 | part | part | HD190228 | NO | NO |
| HD217107 | yes | yes | *HD12661* | *NO* | *NO* | HD114729 | part | part |
| HD162020 | yes | yes | HD150706 | ? | part | HD111232 | ? | part |
| HD130322 | yes | yes | HD59686 | NO | yes | HD2039 | NO | NO |
| HD108147 | yes | yes | HR810 | part | yes | HD136118 | NO | NO |
| *HD38529* | *NO* | *NO* | HD142 | ? | part | HD50554 | NO | NO |
| *55 Cancri* | *yes* | *yes* | HD92788 | NO | part | HD196050 | part | part |
| Gliese 86 | yes | yes | HD28185 | NO | part | HD216437 | ? | part |
| HD195019 | yes | yes | HD142415 | ? | part | HD216435 | NO | NO |
| HD6434 | yes | yes | HD177830 | part | part | HD106252 | NO | NO |
| HD192263 | yes | yes | HD108874 | NO | part | HD23596 | NO | NO |
| *Gliese 876* | *NO* | *NO* | HD4203 | part | part | 14 Her | NO | NO |
| Rho CrB | yes | yes | HD128311 | NO | NO | HD39091 | ? | part |
| *HD74156* | *NO* | *NO* | HD27442 | part | yes | HD72659 | yes** | yes** |
| HD168443 | NO | NO | HD210277 | NO | NO | HD70642 | yes** | yes** |
| HD3651 | ? | yes | HD19994 | part | part | HD33636 | NO | NO |
| HD121504 | yes | yes | HD20367 | ? | part | *eps Eridani* | *NO* | *part* |
| HD178911B | ? | yes | HD114783 | part | part | HD30177 | part | part |
| HD16141 | yes | yes | HD147513 | NO | NO | Gliese 777A | yes** | yes** |
| HD114762 | part | yes | HIP75458 | NO | NO | | | |

**Notes:**
1 The systems are listed by increasing period of the planet with the shortest period, as in the Schneider website, from which planetary orbital elements and further information may be obtained at http://www.obspm.fr/encycl/encycl.html.
2 The $nR_H$ are calculated using the minimum giant masses, though $R_H$ varies slowly, as $m^{1/.3}$.
3 Systems with more than one planet are shown italicised.
4 The column 'now' shows whether an Earth-mass planet could be confined to the HZ within at least the past 1 Gy (excluding the first 0.7 Gy of the main-sequence). If the entry is 'yes' then it could do so almost anywhere in the HZ. If the entry is 'NO' then nowhere in the HZ should offer confinement. If the entry is 'part' then some small proportion of the HZ should offer confinement, for example near its outer boundary for a giant planet not much closer to the star than the inner boundary.
5 The column 'sometime' refers to whether an Earth-mass planet could be confined to the HZ for at least 1 Gy at any time in the main-sequence (again excluding the first 700 Ma).
6 A '?' denotes a star of unknown age, where this is crucial to the evaluation.
7 A '**' denotes those very few cases where the periastron of the giant lies beyond the HZ even at the end of the main-sequence. These are the systems most like the Solar System.